 \documentclass[11pt,preprint2]{aastex}

\slugcomment{Accepted to the Astrophysical Journal}

\shortauthors{Roberts et al.}
\shorttitle{Phase Dependent Spectral Variability of 4U1907+09}

\begin{document}

\title{Phase Dependent Spectral Variability in 4U1907+09}

\author{Mallory S.E. Roberts\altaffilmark{1}, Peter F. Michelson}
\email{mallory@astro.stanford.edu, peterm@leland.stanford.edu}
\affil{Department of Physics, Stanford University,
    Stanford, CA 94305}

\author{Denis A. Leahy}
\email{leahy@iras.ucalgary.ca}
\affil{Dept. of Physics, University of Calgary, Calgary, Alberta, Canada 
T2N 1N4}

\author{Tony A. Hall\altaffilmark{2}, John P. Finley}
\email{hall@egret.sao.arizona.edu, finley@physics.purdue.edu}
\affil{Department of Physics, Purdue University, West Lafayette, IN 47907}

\author{Lynn R. Cominsky}
\email{lynnc@charmian.sonoma.edu}
\affil{Department of Physics and Astronomy, Sonoma State
University, 1801 E. Cotati Ave, Rohnert Park, CA 94928}

\and

\author{Radhika Srinivasan}
\email{rsriniva@mrsc.ucsf.edu}
\affil{Department of Radiology/MRSC, UCSF, San Francisco, Ca.}

\altaffiltext{1}{Current address: McGill University, Physics Department, 
3600 University St., Montreal, Quebec, H3A 2T8}

\altaffiltext{2}{Current address: Iowa State University c/o Smithsonian
Institution, Steward Observatory 933 N. Cherry Ave., Tucson, AZ 85719}

\begin{abstract}
We report on ASCA, RXTE, and archival observations of the high mass 
X-ray binary pulsar 4U1907+09. Spectral measurements of the absorption and flux
were made at all phases of the X-ray pulsar orbit, including the first spectral
measurements of an extended period of low flux during two of the ASCA
observations. We find that a simple spherical wind model can fit the 
time averaged light curve as measured by the RXTE ASM, but does not fit the 
observed changes in the absorption column or account for the
existence of the phase-locked secondary flare. 
An additional model component consisting of a trailing stream can
account for the variations in column depth. However, these models favor a high
inclination angle for the system, suggesting a companion mass more 
consistent with an identification as a Be-star. In 
this case an equatorially enhanced wind and inclined 
neutron star orbit may be a more appropriate interpretation of the data.

\end{abstract}


\section{Introduction}

4U1907+09 was noted as a variable X-ray source during the early
sky surveys with Uhuru \citep{g71}, Ariel V \citep{s76}, OSO-7
\citep{m78}, and HEAO-1 \citep{s80}. The latter allowed Schwartz
et al. (1980) to identify the source with a heavily reddened $m_v \simeq
16.4$ star having a broad $H_{\alpha}$ emission line, which they
assumed was an OB supergiant. A large X-ray outburst observed by
Ariel V during January 1980 indicated regular flux modulation on a
time scale of days. Further analysis of all the Ariel V data
revealed a period of 8.4 days \citep{mr80}. The folded period
profile shows a large primary peak as well as a smaller secondary
maximum $\sim$half a cycle later. Long term observations by Vela
5B seemed to indicate an additional quasi-periodicity at $P\sim41.6$ 
days, which was proposed to be due to precession similar to
that of Her X-1 \citep{pt84}.

In 1983, Tenma observed 4U1907+09 for an entire 8 day cycle, which
resulted in the discovery of double peaked 437.5 s pulsations
\citep{m84}. Analysis of the pulse arrival times allowed an
estimate of the orbital parameters to be made, which confirmed the
8.38 d period as the orbital period. A moderately eccentric
($e\sim0.22$) orbit best fit the data, although a circular orbit
also provided an adequate fit. Makishima et al. also determined a
mass function of $\sim9M_{\sun}$, confirming the massive nature
of the companion. Further observations by EXOSAT \citep{cp87}
implied a mean spin-down rate of $\sim0.23$ s yr$^{-1}$, a
variable column density ($1.5-5.7 \times 10^{22}\ \rm
{cm^{-2}}$), and slightly refined the Tenma orbital elements. The
observed spectra were consistent with a hard power-law ($\alpha
\sim1.2$) with variable low-energy cut-off. Balloon observations
of 4U1907+09 in the 20-100 keV range and analysis of Einstein
archival data are consistent with the EXOSAT spectral observations
\citep{c93}.

Ginga spectral studies suggested a cyclotron absorption feature at
$\sim19$ keV \citep{m95,mak99}, which has been confirmed by a recent
SAX observation \citep{c98}, along with the discovery of the
second harmonic at 39 keV, suggesting a surface magnetic field
strength of $\sim2\times10^{12}$ Gauss. An iron emission line at
$\sim6.4$ keV was also detected, with an equivalent width of
$\sim60$ eV.  A pulse period of $\sim439.5\pm0.6$ s is clearly seen in
the Ginga data during most of the observations where the source had an 
average intensity of 15-20 mCrab.  However 4U1907+09
also exhibited extended periods of weaker intensity ($\sim$3
mCrab) accompanied by a reduced pulsed fraction \citep{mak99}.

Iye (1986) performed the first detailed spectroscopic studies of
the companion star to 4U1907+09. He concluded, in contrast to the
initial assumptions by Schwartz et al. (1980) that the spectral
type of the companion was B2 III -- Ve, and that the two X-ray
flares seen in some orbits were due to the intersection, and
subsequent accretion of material in an extended disk around the
companion by the orbiting neutron star.

\citet{vvv89} next undertook extensive spectroscopic
studies to determine whether the companion to 4U1907+09 is an OB
supergiant or a Be-star. They considered X-ray, optical and
orbital data and suggested most of the measurements support
the supergiant nature of the companion, including reddening and
distance, $H_{\alpha}$ FWHM,  and rotation velocity of the
companion. In addition, the location of 4U1907+09 in the spin-period versus
orbital-period diagram \citep{c84} is inconsistent with the 
trend observed for most Be/X-ray pulsars, suggesting instead a
wind-fed supergiant system. Since
it is difficult to produce a phase-locked secondary flare from the
wind of a supergiant star, \citet{vvv89}
suggested the secondary flare may be spurious in nature.
There is one clear case of a
wind-fed supergiant/X-ray pulsar system, GX 301-2, which has a tendency
to flare twice
per orbit for extended periods of time \citep{pr95}, but the cause of this
is not well understood. 

Recent observations of 4U1907+09 with RXTE showed highly variable flux with
frequent drops to below the detection threshold \citep{isb97}.
There is no evidence for increased absorption during the drops in
their data, suggesting the low flux levels are due to a low
accretion rate. A possible flare that is consistent in phase with
previously observed secondary flares was also seen in the RXTE
data. The 2-30 keV RXTE spectrum is well fit by a cutoff power-law
having a photon index $\alpha \sim1.0-1.3$ with a cutoff energy
of 13.6 keV, and $N_H$ varying from $1-9 \times 10^{22}~ {\rm
cm^{-2}}$, the lowest values occurring during the secondary flare.
No evidence for line emission was found with $\sim60$ eV upper
limit on an iron line equivalent width.

By combining the RXTE data with the Tenma data, in't Zand, Baykal,
and Strohmayer
(1998) obtained a slightly improved orbital fit. This allowed a
good determination of the orbital period (8.3753 $\pm 0.0003$
days). However, the eccentricity is still not well determined,
with a 68\% confidence region of 0.14-0.38. They also reported
the discovery of 18 s oscillations during a 1 hr flare, which may
be the result of the beat frequency between the spin period and an
accretion disk. In addition, the measured spin period
$P_s=440.341^{+0.012}_{-0.017}$ suggests the pulsar has been
steadily spinning down since 1983, with $\dot P_s=+0.225 {\rm s~
yr}^{-1}$.

In this paper we present timing and spectral analysis of four
ASCA observations of 4U1907+09 and spectral analysis of an additional
9 RXTE observations, roughly equally spaced around the
8.4 day orbit. We also report on archival observations of this
source with HEAO A-1 and the RXTE All Sky Monitor. We then compare these 
and all previous observations of 4U1907+09 to a model of accretion from
a spherical wind coming from the companion.

\section{Archival Observations}

\begin{figure}[h!]
\plotone{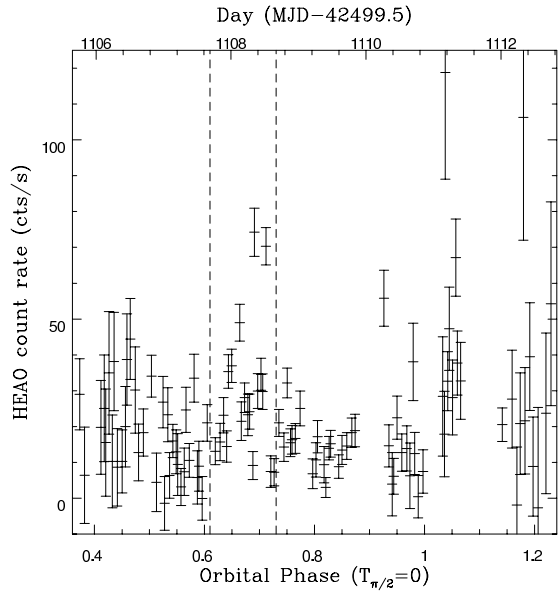}
\caption{\label{HEAOfig} Light curve of 4U1907+09 during the 1978
HEAO A-1 observations. The vertical lines in this and the following figures
represent the 68\% confidence region of the time of periastron. }
\end{figure}

We have investigated archival HEAO A-1 scanning data on 4U1907+09.
HEAO A-1 observed 4U1907+09 in a series of 32 s scans during 1978
April 7 -- 14. The data show orbital behavior consistent with the
Tenma/EXOSAT period with two peaks at orbital phases consistent
with what was seen by Ariel V. The lightcurve from the A-1
observations is shown in Figure~\ref{HEAOfig}.

\begin{figure}[h!]
\plotone{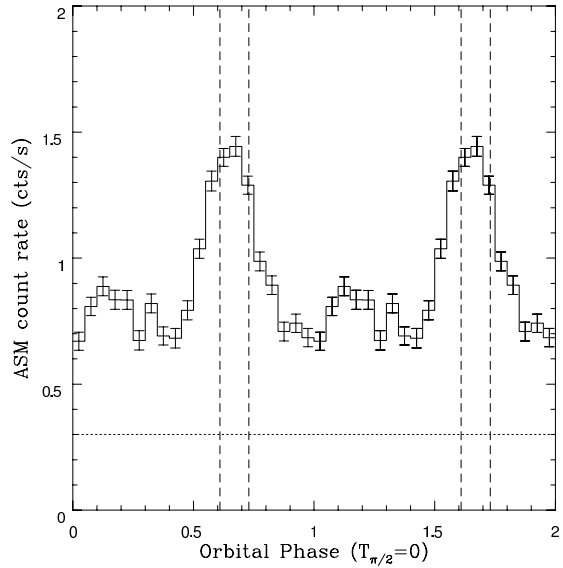}
\caption{\label{ASMorb} RXTE All Sky Monitor
light curve of the 8.3753 day orbital period averaged over the 4 year
data set. The dotted line is the level of positive flux bias assumed
for later model fits.}
\end{figure}

The RXTE All Sky Monitor has been observing this source since it
began regular observations in February, 1996. We report here on
$\sim1400$ days of data (MJD 50136-51549). 
Errors on individual flux measurements
are relatively large, making examinations on time scales of days
unreliable. However, epoch folding of the light curve clearly
shows the orbital modulation. In Figure~\ref{ASMorb} we show the
average orbital light curve for the entire
data set, with the orbital period and epoch $T_{\pi/2}$ from \citet{ibs98}. 
The vertical lines represent the 68\% confidence region 
of the time of periastron.
There is evidence for a secondary flare at a phase of $\sim0.15$. 
We also folded separately the
light curves from the 3 energy channels (1.3-3.0 keV, 3.0-5.0 keV, and
5.0-12.1 keV) of the ASM and found evidence for the secondary flare at a 
consistent phase in each. 
A comparison of the first half of the data
set to the second showed this is a fairly stable feature. 
The ratio of the flare to quiescent flux 
is difficult to estimate since there is known to be a positive flux
bias in the ASM data. In the case of 4U1907+09 this bias is exacerbated by
several nearby sources including GRS 1915+105 and W49B. 

\begin{figure}[h!]
\plotone{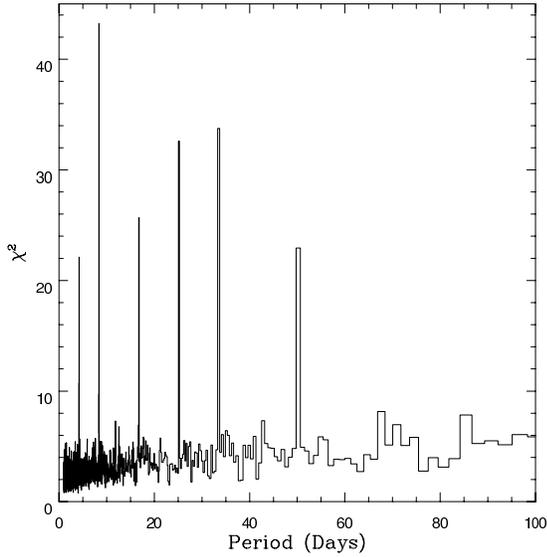}
\caption{\label{ASMper} Period search of the RXTE All Sky Monitor
data showing the 8.4 day orbital period and multiples. }
\end{figure}

We have also performed a period search on the ASM data in order to
search for other periodicities such as the proposed 41
day periodicity \citep{pt84}. We binned the data into 20 phase bins
and calculated $\chi^2_{\nu}$ for a constant model averaging over
phase, as recommended by Collura et al. (1989). 
As Figure~\ref{ASMper} shows, the
8.375 day orbital period and several multiples of it are poorly fit
by a constant model. Although not
evident in the ASM data folding, five times the orbital period would correspond
to 41.9 days
which is consistent with the quasi-periodicity seen in the {\it Vela 5B} data
by  \citet{pt84} and interpreted as an additional,
longer period. We find no evidence for a periodicity in the long-term 
light-curve of
this system which is unrelated to the orbital period.

\section{ASCA Observations}

In August 1996, ASCA made four 10 ks observations of 4U1907+09,
roughly equally spaced throughout the 8.4 d orbit. Observational
parameters are listed in Table~\ref{obstable}.
Extraction regions of 6'
radius were used for the GIS detectors, while a 4' radius was used
for SIS0. A slightly smaller region was used for SIS1 ($\sim3.5'$) 
since the source was nearer the edge of the chip. Since the
source is in the galactic ridge a local estimate of the background
is necessary. Therefore, source free regions of the detectors were
used for extracting background light curves and spectra with a
background rate of $\sim 3\times 10^{-2} {\rm cts/s}$ in each detector.

\subsection{Temporal Analysis}
\begin{figure}[h!]
\plotone{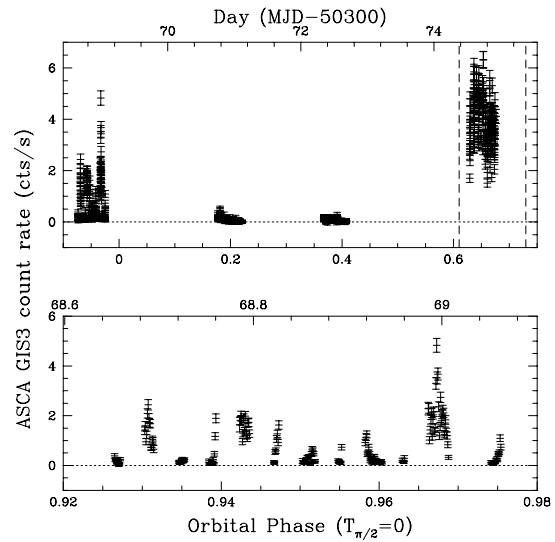}
\caption{\label{ASCAglc}  The ASCA GIS light curve of the four
observations of 4U1907+09 (upper panel) and of just Aobs1 (lower panel). }
\end{figure}

\begin{figure}[h!]
\plotone{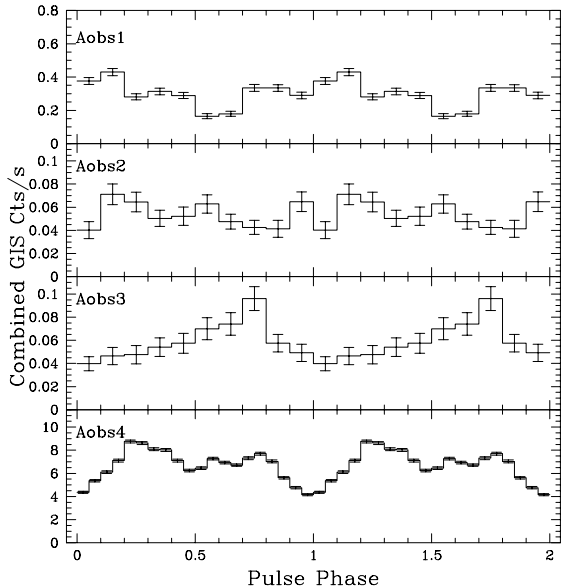}
\caption{\label{ASCApulse} Pulse profiles from the ASCA observations. The
photon arrival times are corrected for the pulsar orbit, and folded from a
common epoch using the best fit period of $440.35$s from epoch folding obs4.
The error in the period results in a relative phase uncertainty
between obs1 and obs4 of $\sim0.15$.  }
\end{figure}

The background subtracted GIS3 light curve for all 4 observations (Aobs1-4)
is shown in Figure~\ref{ASCAglc}. Aobs1 consists of short flaring episodes
with times of low flux in between where the count rate dropped by
a factor of $\sim10$, reminiscent of the ``dipping" seen in the
earlier RXTE observations \citep{isb97}. Aobs2
and Aobs3 show the source in a very low state, a factor of $\sim50$ below
the average flux in Aobs4 and about a factor of $\sim4$ below the low
flux states in Aobs1. Since Aobs4 has the best statistics and is
relatively constant on time scales longer than the pulse period,
we used it to find the best pulse period of $440.35\pm0.06$ sec (after
correcting the arrival times for the orbital motion of the pulsar, assuming the 
orbital parameters of \citet{ibs98}) by epoch
folding the light curve. We then folded each of the
observations at this period and show these in
Figure~\ref{ASCApulse}. The pulses seem to change shape and 
become weaker during the low states (Aobs2 and Aobs3).

\subsection{Orbital Phase Spectroscopy}

Using Xselect, spectra were extracted from each of the four
detectors for each observation. These were analyzed using XSPEC
and fit to an absorbed power law model, adding narrow ($\sigma =
0$) Gaussian lines if they significantly improved the fit. Aobs2
and Aobs3 had very low statistics so the two GIS detectors were
combined into one data set, as were the two from the SIS
detectors. In order to use all available photons, SIS Bright mode
data were used. The results are shown in Table~\ref{spectablea}. It can
immediately be seen that the lower states (Aobs2 and Aobs3) show no
evidence of increased absorption. The lowest absorption occurs
during Aobs4, the brightest observation, and is a factor of 2 lower
than the high absorption seen during Aobs1. The spectral slope
changes by a small, but statistically significant, amount. Weak
iron line emission is present during Aobs4 and may be present
during the other observations, although the statistics were not
good enough to give more than upper limits at the 90\% confidence level.

\placetable{spectablea}

\begin{figure}[h!]
\plotone{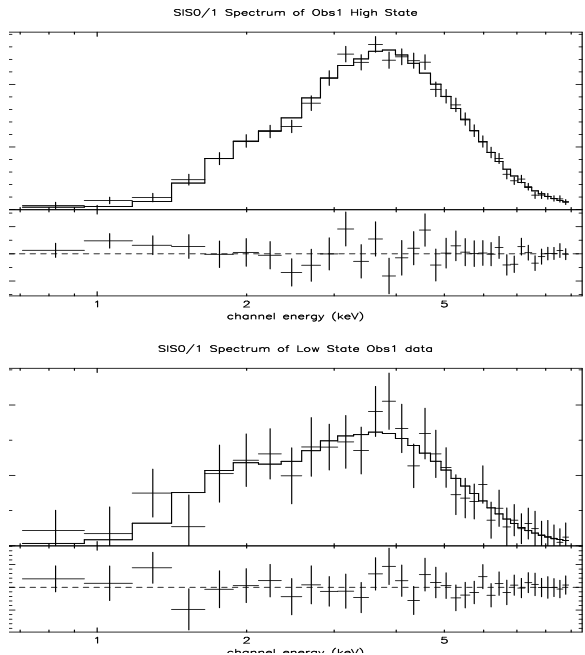}
\caption{\label{ASCAspec}  High state and low state SIS energy spectra for
Aobs1. }
\end{figure}

In order to test whether the drops in flux during Aobs1 are due to
an increase of column depth, we extracted spectra separately from
the flaring and the low portions of the data set. As shown in
Figure~\ref{ASCAspec},  there is no evidence of increased
absorption during dips. In fact, there is evidence that the absorption column
$decreases$ during the dips, although the statistics on the low
flux are such that the data are barely consistent within the 90\%
multi-parameter confidence region. There is no evidence of a
significant change in photon index. If we assume the photon index remains
constant, the change in absorption column is $\sim2\times
10^{22} {\rm cm^{-2}}$. In the SIS data during the periods of low
flux there appears to be an emission feature near 3.8 keV, which
is evident in both SIS0 and SIS1, although not apparent in the GIS
data. In the GIS data the fit could be significantly improved by an
inclusion of a 0 width Gaussian line at 2.2 keV. This feature
appears in several of the data sets, although since it is near the
known bump in the response due to Gold in the telescopes, this
feature should be viewed with caution.

\begin{figure}[h!]
\plotone{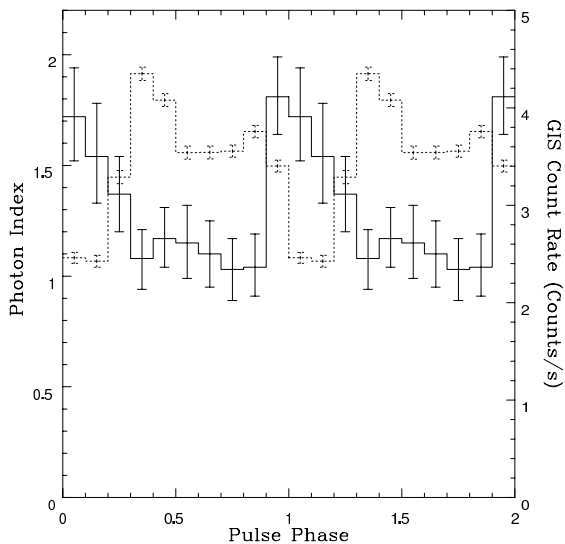}
\caption{\label{pulspec} The photon spectral index as a function
of pulse phase (arbitrary epoch) during Aobs4.
Errors represent the 90\% confidence region.
The dotted histogram is the GIS count rate. }
\end{figure}

 During Aobs4 the source was only varying
by a factor of 2 or so, mostly due to the pulsations, and had a
fairly high count rate allowing tight constraints on $N_H$ and
the photon index. The fits are significantly improved by the
inclusion of an iron line at $\sim6.4$ keV. The count rates were
sufficiently high during this observation to
extract spectra as a function of pulse phase 
using the best fit pulse period. There is no
strong evidence for variability in $N_H$. A strong
anti-correlation between photon index and intensity is evident which we
show in Figure~\ref{pulspec}. There also seems to be a
concentration of iron line emission in the pulse phase 0-0.1 spectrum.
However, the uncertainties in the fit line strength at the
other phases preclude a definitive measurement of line strength variability.

\section{RXTE Observations}

RXTE observed 4U1907+09 for a second series of observations during December
1996. A total of 9 observations (Robs1-9), each approximately 2 ksec in
duration, were made spaced roughly one day apart in order to evenly sample
the 8.4 day orbit. Data from the Standard2 observational mode were analyzed 
using FTOOLS 4.2. Combined spectra were extracted from all 5 PCUs (except
obs1, which only had PCUs 0-3 on during the observation) and all layers. 
Model background spectra were then constructed using the background models
provided by the RXTE Science Operations Facility, which take into account the particle induced background. However, they do not take into account
the X-ray background from the galactic plane. To construct an accurate
background model for 4U1907+09, it is necessary to take into account 
contributions from both
the Galactic ridge and the nearby supernova remnant W49B. 
During two of the observations, Robs4 and Robs5, the source was in the ``dip"
state. The model background was subtracted from these observations, and their
spectra were examined. The count rate of $\sim 18$cts/s is comparable to what
is observed by RXTE for the diffuse emission from the Galactic ridge
\citep{vm98}. We created a crude model of the Galactic ridge 
plus SNR W49B emission consisting of two 
thermal 
bremsstrahlung sources with fixed temperatures of $1.8$keV and $\sim10$ keV, 
moderate ($1.6\times 10^{22} {\rm cm}^{-2}$) absorption, and a $\sim6.7$ keV
gaussian lines with only the normalizations allowed to vary. These parameter
values were taken from \citet{k89} for this position in the 
galactic ridge and \citet{s85} for W49B. This model adequately fit the
data for Robs4 and Robs5 with normalizations that could reasonably
be expected if the source 
had dropped
to flux levels consistent with the low states seen in the ASCA data and the
spectra were dominated by emission from the Galactic ridge and the nearby 
thermal supernova remnant. Therefore, these observations were used to make a 
background spectrum for use in spectral fits of the other observations in the
following way. Using the background models available from the RXTE GOF, 
we generated background spectra for all 9 observations. These were
directly subtracted from the observation spectra using the mathpha 
FTOOL to create the files to be used in the fitting.
The background subtracted pha files of Robs4 and Robs5 were then added 
together to create a local X-ray background for 4U1907+09 which
was used as the background file in XSPEC. 

All the spectra from the remaining seven observations were modelled using
an absorbed power-law with a high-energy cutoff in XSPEC. The data were fit 
in the energy range of 2.5 - 20 keV, and the results are listed in 
Table~\ref{spectabler}. The measured cut-offs 
are consistent with the broad-band spectral fits of Beppo-Sax \citep{c98}. 
We also obtained the archival RXTE data from February 1996 first analyzed 
by \citet{isb97} to fit for average fluxes and absorption in the 2.5-10 keV
band for inclusion in the modelling below.  
 
\begin{figure}[h!]
\plotone{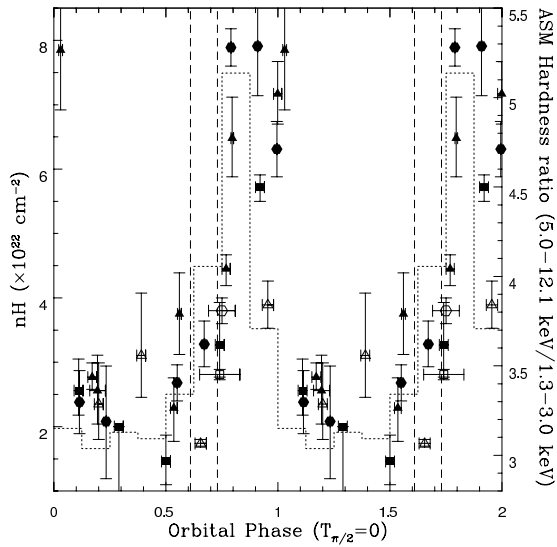}
\caption{\label{ORBvar} The multi-mission measurement of nH as a
function
of orbital phase. Measurements are from ASCA (open triangles), Feb. 1996
RXTE (filled triangles), SAX (open square), EXOSAT (filled square), Tenma
(open hex), and Dec. 1996 RXTE (filled hex). The histogram is the ASM hardness
ratio of the countrates from channel 3 over channel 1.}
\end{figure}

\section{Modelling of Orbital Spectral Variations}

\begin{figure}[h!]
\plotone{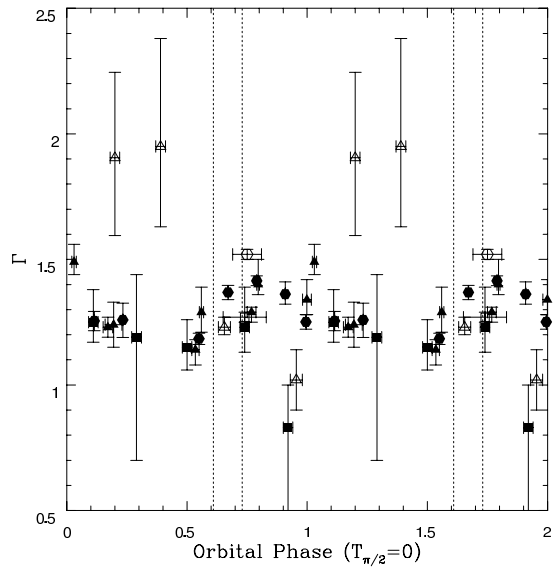}
\caption{\label{ORBsi} The multi-mission measurement of the photon index as a 
function of orbital phase.}
\end{figure}

\begin{figure}[h!]
\plotone{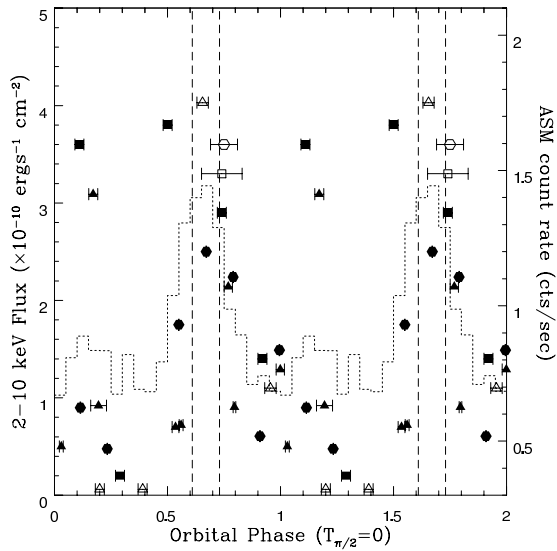}
\caption{\label{ORBflux} The multi-mission measurement of the 2-10 keV
flux as a function of orbital phase. The histogram is the folded lightcurve
from the ASM, with the y axis starting at the assumed flux bias of 0.3 cts/sec.
}
\end{figure}

In Figure~\ref{ORBvar} we show measurements of nH as a function of orbital
phase for spectral measurements from ASCA and RXTE (this work), EXOSAT 
\citep{cp87}, Tenma \citep{m84}, and SAX \citep{c98}. As a consistency check,
we also plot the ASM hardness ratio of (5.0-12.1 keV cts/s)/(1.3-3.0 keV cts/s)
as a function of phase (dotted line).  Since the spectral slope does not vary
greatly, we expect this hardness ratio to change mostly due to column depth.
The vertical lines represent the 68\% confidence region of the time of 
periastron. We see that the  absorption begins
to increase towards the end of the primary flare stays high until right
before the secondary flare, where it drops again. In Figure~\ref{ORBsi}
we plot the photon spectral index as a function of orbital phase. Small
but significant changes are observed. In Figure~\ref{ORBflux}
we plot the 2-10 keV flux measurements from these same observations versus 
orbital phase with the ASM lightcurve overplotted. The secondary flare
is twice seen to be nearly as bright as the primary during these observations. 
One of the RXTE 
observations seen at an overlapping phase with the ASCA Aobs2 is still
seen to be moderately bright ($F_X\sim10^{-10}$ ergs ${\rm cm}^{-2}$
${\rm s}^{-1}$) although dipping activity is seen during the observation. 
Therefore, extended periods of low flux as seen in the ASCA Aobs2 and Aobs3 
probably do not occur every orbit. In general there is a positive flux bias 
of $\sim 0.1$ cts/s in ASM data. However, in crowded
regions of the Galactic plane, such as around 4U1907+09, there are likely
to be further systematic biases.  By comparing the folded ASM light
curve to the observed multi-mission fluxes we estimate a
positive flux bias of $\sim0.2-0.6$ cts/s. 
We will assume a bias of $0.3$ for the analysis below.

In this section we model the primary flare and absorption
as accretion from a spherical stellar wind with the addition 
of an absorbing gas stream. This model, first applied to Tenma observations 
of GX 301-2, is detailed in Leahy (1991).
We first attempt to model the average ASM lightcurve and multi-mission 
absorption data as 
accretion from a 
simple spherical stellar wind with velocity profile:

$$v_{wind}=v_{\infty}(1-R_c/r)^{\beta}$$

\noindent
where $v_{\infty}$ is the terminal velocity, assumed to be 3 times the
escape velocity, and $R_c$ is the radius of the companion. The Castor, Abbott,
and Klein (1975) theoretical wind model has $\beta=0.5$, but observations of
OB stars \citep{l88,b99} suggest that $\beta\sim0.7-2$ depending
on spectral type. We assume $\beta=1$, consistent
with the assumption of an OB supergiant.  In the case of a
lower mass Be-type star, models of the wind structure are more
complex and uncertain. 
However, the velocity profile of the polar wind from a Be-star is similar
to that of an OB supergiant \citep{app94}, and the profile of the equatorial wind 
probably has a similar functional form but lower terminal velocity. If the companion is a Be-star, it is likely
that the spin axis is not aligned with the orbital axis (see discussion) and
so emission dominated by accretion from the polar wind over much of the orbit 
may not be an unreasonable assumption. 

We assume the orbital period $P=8.3753$d, 
longitude of periastron (i.e. the observer's viewing angle) $\omega=330\deg$, 
and periastron phase of the best fit model from
\citet{ibs98} and fit the ASM light curve for a range of inclination
angles by adjusting the eccentricity and the parameter 
$x=\dot M_{20}/d^2_3$, where $\dot M_{20}$ is the mass loss rate in
units of $10^{20} {\rm grams/sec}$ and $d_3$ is the distance in 
units of 3 kpc. 
We choose as our canonical values $i=47.9$, which implies a companion
mass $M_c=24 M_{\odot}$ given the mass function derived from the 
orbital fits assuming a neutron star mass of $1.4 M_{\odot}$, and 
$i=75.7^{\circ}$ for a companion mass $M_c=12 M_{\odot}$ more in line with
a Be-star. 

\begin{figure}[h!]
\plotone{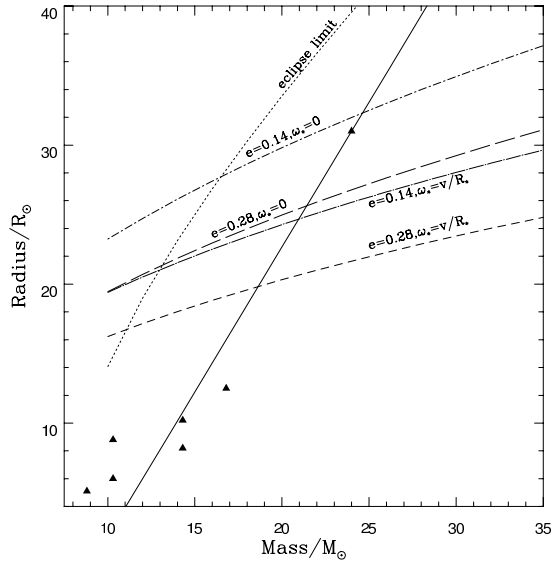}
\caption{\label{ecl} Assumed Mass/Radius relationship
(solid line) compared to eclipse limit and periastron ``Roche Lobe" radii
assuming different eccentricities and stellar rotation rates in the
orbital plane ($\Omega_*$). Points are various stellar values for main
sequence, giant, and supergiant B0-B3 stars \citep{ud82}}
\end{figure}

For the companion radius we follow \citet{vvv89} and assume 
$R_c=31R_{\odot}$ for a $24 M_{\odot}$ star. 
However, a star of this radius would be larger than  its 
Roche lobe at periastron, as defined in Brown and Boyle (1984),
assuming the best fit eccentricity of $e=0.28$ \citep{ibs98}, a rotational velocity
$v_{sini}=90$ km/s \citep{vvv89}, and alignment of the rotational and
orbital axes.  In Figure~\ref{ecl} we plot the Roche lobe radius 
as a function of companion mass as well as the limits on radius
inferred from the lack of eclipses. All supergiant companions
would greatly overflow their Roche lobes and would have to be
excluded if we assume this as a radius limit. However, a non-rotating
(or non-aligned) star has a greater Roche lobe radius, as would
a system where the eccentricity is less. 
For the $12M_{\odot}$ star, we
use the \citet{i86} value of 
$R_c=6R_{\odot}$ for a Be-star companion.
For intermediate inclinations and masses, the apropriate 
radius is dependent on spectral type. For supergiants, 
the radius actually increases with decreased mass, which would
exacerbate the Roche lobe  problem. However, a type
Ib, III, or V are all smaller than our
canonical  Ia supergiant star.  In Figure~\ref{ecl} we plot 
radii for a number of early B stars \citep{ud82}.
A linearly interpolation between our two canonical stars
seems to provide a reasonable mass-radius relationship
for modelling purposes.

\begin{figure}[h!]
\plotone{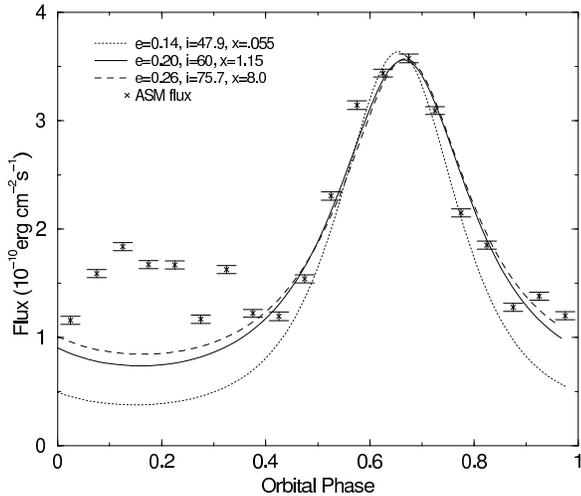}
\caption{\label{LUMmod} Spherical wind model fits to ASM
lightcurve assuming inclination angles of $47.9^{\circ}$
($M_c=24M_{\odot}$), $60^{\circ}$ ($M_c=16M_{\odot}$),  and $75.7^{\circ}$
($M_c=12M_{\odot}$).
}
\end{figure}

In Figure~\ref{LUMmod}
we show the range of model curves compared to the ASM light curve. 
To fit the higher mass companion to the shape of the
primary flare an eccentricity $e \la 0.14$ is required (which
would also allow the star to underfill its Roche lobe at periastron). 
For the lower mass companion the best fit to the observed
primary flare peak is $e=0.26$,
more in line with the best fit eccentricity from the pulse arrival analysis.
In no case can a secondary flare be produced from this model.

\begin{figure}[h!]
\plotone{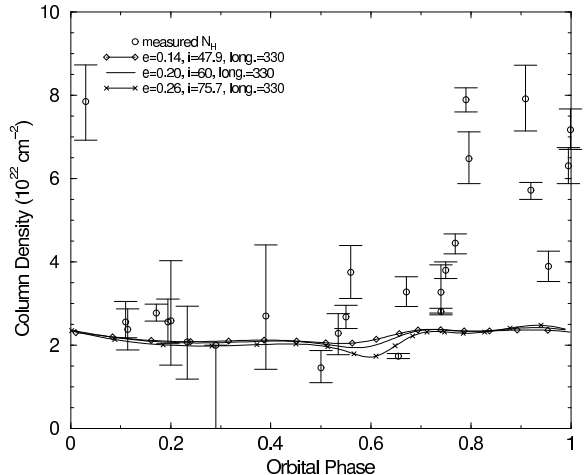}
\caption{\label{MODnh} Spherical wind model absorption compared with
multi-mission spectral fits.
}
\end{figure}

This luminosity model implies a varying absorption 
as a function of orbital phase, with the overall normalization being only 
a function of $\dot M$. Using the parameters given above and an assumed
interstellar absorption of $nH = 1.5 \times 10^{22} {\rm cm}^{-2}$, 
we find that 
$\dot M_{20}=1-1.5$ best fit the absorption data over our range of inclination
angles. This is somewhat low for a B supergiant, but quite high for a Be-star
\citep{app94}.
In order to improve the fit, we also vary the 
the longitude of periastron  since that is poorly constrained by the pulse
arrival time analysis of \citet{ibs98}. 
In Figure~\ref{MODnh} we plot the results compared to the multi-mission data.
Neither the phase or the 
amplitude of the large increase in absorption after
periastron can be produced from this simple spherical wind model.

We can use the $\dot M$ from the absorption fits and the parameter $x$ from
the luminosity model to estimate distances to the system. 
For the $24M_{\odot}$ model, we derive a distance $d\sim4.3$ kpc, while 
for the $16M_{\odot}$ and $12M_{\odot}$ models, we derive $d\sim1.0$ kpc and
$d\sim0.4$ kpc respectively. The first is in line with the distance estimates
of 5.9 to 2.4 kpc for a supergiant companion from \citet{vvv89}, and the latter
are consistent with their estimates of 1.5 to 0.6 kpc for a giant. Note that
the minimum measured $nH\sim1.5\times 10^{22} {\rm cm}^{-2}$ means the
assumed interstellar component of the absorption is an upper limit.
Thus the derived values of $\dot M$, and hence the inferred distances,
should be considered lower limits.

\begin{figure}[h!]
\plotone{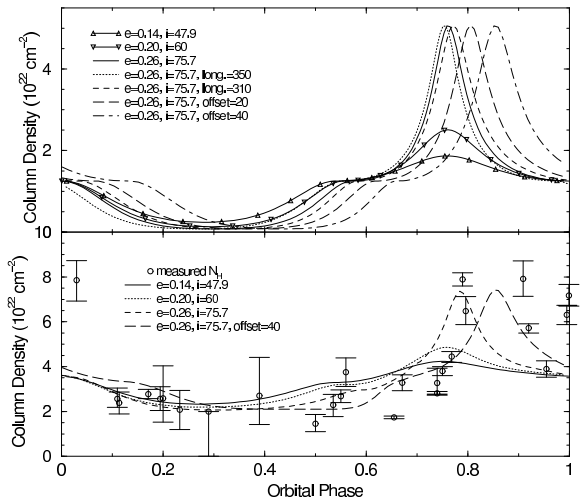}
\caption{\label{WAKEmod} Absorption due to a trailing gas stream.
The top panel shows the orbital variation of absorption due to the stream
for various parameters. The bottom panel shows the wind plus stream
absorption model compared to the multi-mission measurements of nH.
}
\end{figure}

To fit the post-periastron absorption, we added a linear trailing gas stream to
the above model (see \citet{l91} for details). Such a stream can be interpreted as arising from the wake
caused by the passage of the neutron star through the wind, or from an accretion
stream coming from the companion (although this latter interpretation would
require modification of the luminosity model).  The stream is modelled as
a density enhancement with a gaussian profile perpendicular to the stream line,
an exponential cutoff on the front side and no cutoff in the downstream
direction. The total contribution to the absorption  is then obtained by
integrating along the line of site.
For a given orbital phase, longitude of periastron, and orbital inclination
, the amplitude of the 
absorption enhancement of the trailing gas stream is completely determined by the product 
$n_o\sigma\sim10^{22} {\rm cm}^{-2}$, 
where $n_o$ is the central density and $\sigma$ the gaussian width. The
additional absorption due to the stream is shown in the top of 
Figure~\ref{WAKEmod}. The phase of the absorption peak is somewhat earlier
than indicated by the data, so we also tried offsetting the stream line towards
the companion star by angles of $20^{\circ}$ and $40^{\circ}$. The bottom
panel of Figure~\ref{WAKEmod} shows how the combined wind plus stream 
model compares to the absorption data for the various cases. We find that 
an adequate fit can be obtained with the larger inclination angles and
an offset stream. A downstream cutoff of the stream would tend to broaden the
peak somewhat, which might further improve the fit.  

\section{Discussion}

If the eccentricity of $e\ga0.2$ of the best fit orbital parameters 
from pulse arrival time analyses is correct, the primary flare
seen in the overall light
curve can be fit using a model of accretion from a spherical wind off of
a star of mass $M_c \la 16 M_{\odot}$. However, the absorption measurements and
the persistence of the secondary flare require modifications to this
model. We have shown that a trailing accretion stream moderately offset
towards the companion can account for the extra absorption
component, although not the secondary flare. Such an offset stream may be 
indicated by the high resolution two-dimensional hydrodynamical simulations
of \citet{blt97}. The ``flip-flop" instability seen in their simulations 
modifies the direction of the downstream flow so that it switches back 
and forth resulting in a wake that could be characterized by two offset
streams, one of which could provide the enhanced absorption seen in the data
(see Fig. 2 of \citet{blt97}). This instability can in principle also explain the
``dipping" or ``flaring" behavior seen in Aobs1 and in the RXTE observations of 
\citet{isb97}, and may even be compatible with a drop in absorption during
low accretion phases. 
However, this instability may just be an artifact of two-dimensional
simulations, and whether similar affects occur in a three-dimensional
flow is an open question. Recent 3D simulations by Ruffert (1999) do exhibit
active unstable phases, although the changes in accretion rate tend to 
be smaller and the downstream density profile is unclear.

The spherical wind accretion models favor a high
inclination angle, implying a companion mass more in line with a Be-star than
a supergiant. In addition, if the companion mass is high, the models  imply an eccentricity near the low end of the
acceptable range. This lower eccentricity is also required to keep 
a massive supergiant from greatly overflowing its Roche lobe near periastron. 
Even with a smaller eccentricity, it is likely that periodic Roche lobe
overflow \citep{bb84} or tidal stripping \citep{l97} would be the dominant 
mass transfer mechanism. In this case, the observed accretion rate and orbital variation
may be somewhat low, unless long-term storage of material by an accretion disk
could keep the accretion rate low except for occasional large outbursts like
the one seen by Ariel V in 1980 \citep{mr80}. 
If the
best fit eccentricity value of $e\sim0.28$ is correct, it would seem to exclude the 
accretion from a supergiant wind model. 
This suggests that an alternative model of the accretion
perhaps should be considered. One persistent possibility
is an equatorially enhanced wind with the neutron star in an inclined orbit,
as suggested by Iye(1986).
In this case, the neutron star passes behind an extended matter disk around
the companion after periastron and comes back through after apastron. 
While tidal forces would tend to align the neutron star orbital axis with that
of the companion's spin axis, the timescale for this is similar to the
circularization timescale, and accretion processes might counteract
the tidal forces somewhat \citep{mw83}. 
Given that the orbit is eccentric, we do not view a somewhat inclined orbital 
axis as being unreasonable.

In the equatorially enhanced wind scenario,
the flares are fed by matter gathered by the neutron star as it
passes through the matter envelope. If the outflow speed is 
somewhat slower near
the equator, the size of this envelope might be
limited by the neutron star's orbit. Such truncation is indicated from studies
of the H$\alpha$ emission line in several Be/X-ray binary systems
\citep{z00}. The matter is depleted by the
periastron passage of the neutron star, so significantly less material
is available further out at the apastron radius. If sufficiently depleted,
there may be no secondary flare at all. Hence we get an intermittent but
phase locked secondary flare. 

The major difficulty to this picture is the ``virtually certain" classification of the
companion as a supergiant by \citet{vvv89}. This is based on a collection
of observed properties of the system, with some favoring a Be-star, and 
some
favoring a supergiant. Two factors seem to greatly favor the
supergiant interpretation. First is the color excess $E_{B-V}=3.4\pm0.2$,
which is quite high for the inferred distance of $\la 1.5$ kpc of a Be-star. 
However, the variation in absorption column with orbital phase suggests that 
much of this could be local. Indeed, the average excess 
determined from interstellar
absorption lines is only $E_{B-V}=2.4\pm0.5$ or $E_{B-V}=2.7\pm0.6$, depending
on which coefficients for the ratio of equivalent
width to color excess used \citep{vvv89}. These lower values are more in line with the 
inferred
interstellar X-ray absorption of $nH\la 1.5\times 10^{22} {\rm cm}^{-2}$.
The independent distance estimate \citet{vvv89} obtain 
from the interstellar sodium lines
of $d=2.1\pm0.6$ is  between the supergiant and Be-star distances, and
consistent with both given the error. 

The second factor strongly  favoring the supergiant interpretation is 
the large FWHM$\sim$700 km/s of the H$\alpha$ line combined with the
relatively low projected 
rotational velocity $v_{sini}\sim 90$kms. While neither
value is exceptionally unusual for a Be-star, the combination is \citep{d86}.
The inclination angle suggested by the observed projected rotational velocity
is $i\sim 15^{\circ}$ assuming typical Be-star rotational velocities, 
as opposed to the inferred orbital inclination $i\sim 60^{\circ}-75^{\circ}$
of the neutron star with  a giant or main sequence star companion, 
consistent with the inclined orbit scenario.
However, the large value of the FWHM suggests the outflowing disk is seen
nearly edge on. The circumstellar envelope can and would be affected 
by the neutron star, as evidenced by the large-scale perturbations observed
in the $H\alpha$ profiles of several Be/X-ray binaries \citep{n98}. Since 
the orbital period of 4U1907+09 is shorter than any of the known
Be/X-ray binaries (except A0538-66, which is not well observed), it would
be expected to have a much greater perturbing effect. It is interesting to note
that the projected orbital velocity of a particle in the orbital plane near the 
Roche lobe radius of 
a lower mass star in 4U1907+09 is roughly half the observed H$\alpha$ FWHM. 
If the gravitational attraction of the neutron star could warp the
equatorial outflow of the companion such that it would be nearly
co-planar to the neutron star orbit at the Roche lobe radius, the observed
FWHM might result. Hydrodynamical simulations should be able to determine if
such a scenario is feasible. 

The relatively long periods of low accretion rate seen in Aobs2 and Aobs3
are somewhat puzzling.
Inhomogeneities in the wind from the companion could possibly account
for these variations, although the apparent duration of 
several days during the ASCA observations would argue against this.
If the accretion is modified by a transient accretion disk, 
instabilities in the accretion disk could account for both the sudden changes
in accretion rate seen in Aobs1 and the longer periods of quiescence seen in
Aobs2 and  Aobs3. 
The consistent spin-up of the neutron star and the 18 sec. 
oscillations seen by RXTE \citep{isb97} argue for the existence of an
accretion disk.
The on-off behavior seen by both RXTE and ASCA are suggestive
of magnetically inhibited accretion from a disk whose
inner edge is near the corotation radius. The residual emission
during the off periods may then be a result of accretion on to
the magnetosphere or a bow shock. However, the slow
spin period and correspondingly large corotation radius makes this
unlikely.

\section{Comparison with GX301-2}

4U1907+09 is similar in many respects to the wind fed Supergiant/X-ray 
pulsar binary system GX301-2. Both have eccentric orbits and occupy
a similar location on the spin-period versus orbital-period diagram
\citep{c86} (for GX 301-2, $e\sim0.47, P_{spin}\sim680 {\rm s}, P_{orb}\sim
41.5 {\rm d}$, \citet{k97}). The 
absorption profile and primary orbital flux
peak of both systems have been successfully fit by a spherical wind plus gas 
stream model \citep{l91}. GX 301-2 also has extended periods where there is
a secondary orbital flare at a consistent phase that cannot be
fit by wind models. An equatorialy enhanced outflow has also been suggested
for this system and models of such an outflow can grossly fit
the orbital light curve \citep{k97}.
Both systems have companions with similar temperatures (spectral type B2) 
and projected rotational velocities ($v\sin i \sim 70$ for GX 301-2,
\citet{p80}).

However, the companion star of GX 301-2,
Wray 977, is unambiguously identified as a supergiant by the strength
of various lines in the blue portion of the spectrum \citep{p80}.
The observations of \citet{i86} and \citet{vvv89} did not extend into
the blue portion of the spectrum, so such diagnostics are not available
for 4U 1907+09. The H$\beta$ emission line of Wray 977 has a P-Cygni
profile \citep{p80}, clear evidence of an expanding atmosphere. 
No such profiles have been seen in 4U 1907+09. 

The local absorption column and iron line equivalent width
of the X-ray spectrum of GX 301-2
are both more than an order of magnitude stronger than in
4U1907+09. This implies the mass loss from Wray 977 is much greater than
from the companion of 4U1907+09, despite the similarity in the proposed
spectral classes. While enhanced mass loss in these systems could be expected 
due to gravitational interaction with a neutron star,
the orbit of GX 301-2
is significantly wider, and there is no question that the companion
remains inside its Roche lobe (although at periastron it can come close,
\citet{l91}). If both were supergiants, the naive expectation 
would be greater mass loss from the companion of 4U1907+09 due
to transient Roche lobe overflow \citep{bb84} or tidal stripping
\citet{l97}.

The changes in spectral index with pulse phase of both systems 
show a softening of the spectrum near the pulse minimum in the
1-10 keV band \citep{s96,lm90}. The iron line equivalent width
in GX 301-2 varies with pulse phase as well, with a peak near the
pulse minimum. The possible iron line enhancement near the pulse minimum
in the 4U1907+09 {\it ASCA} data, if real, would be consistent.
These similarities in the phase-resolved spectra suggest the 
accretion columns are similar. This is not surprising, since the 
spin periods and surface magnetic field strengths (as inferred
from their cyclotron absorption lines) are similar. 

\section{Conclusions}

Observations of the variation of the spectra of 4U1907+07 with orbital 
phase are consistent with a simple spherical wind accreting onto
the neutron star from an OB supergiant companion with excess
absorption caused by a trailing accretion wake. However, the implied
inclination angle for such a model suggests a relatively low mass 
($\la16M_{\odot}$) for the companion. The excess of 
absorption column seen after periastron might also be interpreted
as the neutron star passing through
an equatorially enhanced matter envelope in an inclined orbit. 
Observations of the 
secondary flare near apastron by RXTE and HEAO A-1, as well as the average orbital 
light curve from the RXTE ASM, further support the hypothesis that this 
flaring has been occurring at a consistent phase but with variable amplitude
for 20 years. The 
measurements of the low states by ASCA suggest that the accretion 
rate can change by 1-2 orders of magnitude over a short time period, 
perhaps due to a transient accretion disk. The most critical components
to a better understanding of this system are more secure measurements
of the orbital eccentricity and periastron phase and improved optical 
measurements of the companion. A high resolution optical spectrum at 
shorter wavelengths and  
orbital radial velocity measurements could unambiguously determine the 
companion type. 

\acknowledgments 

We thank J. G. Jernigan for useful discussions as
to the nature of the companion star in this system and the
anonymous referee for pointing out the importance of the Roche lobe radius. 
All Sky Monitor results
were provided by the ASM/RXTE teams at MIT and at the RXTE SOF and GOF at
NASA's GSFC. 
This work was supported in
part by NASA Guest Observer grants NAG 5-2948 and NAG 5-3332.

\clearpage
\begin{deluxetable}{lccccc}
\footnotesize
\tablecaption{ASCA Observation Parameters
\label{obstable}} \tablewidth{0pt}
\tablehead{\colhead{Observation}& \colhead{Sequence ID}& \colhead{Date} &
\colhead{Orbital Phase}   & \colhead{On-Time}   & \colhead{Mean
GIS rate}\\ &&MJD-50300&$T_{\pi/2}=0$&ks&Cts/s } \startdata Aobs 1 &44020040
& 68.65-69.06 & 0.93-0.98 & 10.2 & $0.702 \pm 0.008$ \\ Aobs 2 &44020010&
70.75-71.12 & 0.18-0.22 & 10.1 & $0.067\pm0.005$\\ Aobs 3 &44020020&
72.34-72.53& 0.37-0.39 & 9.3 & $0.064 \pm 0.005 $ \\ Aobs 4 &44020030&
74.54-74.94 & 0.63-0.68 & 9.9 & $3.447 \pm 0.015$ \\
\enddata
\end{deluxetable}


\begin{deluxetable}{lcccccc}
\footnotesize
\tablecaption{ASCA Spectral Fits \label{spectablea}}
\tablewidth{0pt} \tablehead{\colhead{Observation}& \colhead{$n_H$}
& \colhead{$\Gamma$}   & \colhead{line E}& \colhead{line
EqW}&\colhead{$F_{2-10keV}$}&\colhead{$\chi^2_{\nu}$(dof)} \\ 
\colhead{} &  \colhead{$\times 10^{22} {\rm cm}^{-2}$} & \colhead{} & 
\colhead{keV} & \colhead{eV} &\colhead{$\times 10^{-10}$ ergs ${\rm cm}^{-2}
{\rm s}^{-1}$}  &  }
\startdata 
\multicolumn{7}{c}{Fits from GIS2 + GIS3} \\
Aobs 1 &
$3.89^{+0.37}_{-0.36}$ & $1.02^{+0.12}_{-0.12}$ & $6.53^{+0.32}_{-0.52}$ & 
$76\pm33$ & $1.11\pm0.02$ & 0.58(38) \\ 
Aobs1 hi & $4.18^{+0.41}_{-0.38}$ & $1.05^{+0.13}_{-0.12}$
 & $6.45^{+0.31}_{-0.22}$ & $82\pm33$ & $1.73\pm0.02$  & 0.50(38) \\ 
Aobs1 lo & $2.12^{+1.68}_{-1.12}$ & $0.92^{+0.67}_{-0.53}$ &  &  &
$0.23\pm 0.02$  & 0.24(40)\\ 
Aobs 2 & $2.59^{+1.44}_{-1.07}$ & $2.07^{+0.81}_{-0.65}$ & & & 
$0.066\pm0.005$ & 0.25(40) \\ 
Aobs 3 & $2.70^{+1.71}_{-1.28}$ & $1.93^{+0.88}_{-0.73}$ & & & 
$0.066\pm0.005$ & 0.19(40) \\ 
Aobs 4 & $1.74^{+0.06}_{-0.06}$ & $1.23^{+0.04}_{-0.03}$ & 
$6.53^{+0.11}_{-0.11}$ & $79\pm16$ & $4.20\pm0.02$ & 1.71(38)\\
\multicolumn{7}{c}{Fits from SIS0 + SIS1}  \\
Aobs 1 & $3.83^{+0.39}_{-0.36}$ & $1.03^{+0.14}_{-0.14}$ & 6.34 (frozen) & 
$< 57.1$ & $0.87\pm0.01$ & 0.53(289) \\ 
Aobs1 hi & $4.01^{+0.42}_{-0.39}$ & $1.04^{+0.14}_{-0.15}$ & 6.34 (frozen) & 
$< 65.8$ & $1.42\pm0.02$  & 0.51(289) \\
Aobs1 lo & $2.47^{+1.72}_{-1.20}$ & $0.96^{+0.69}_{-0.60}$ &  &  &
$0.18\pm 0.01$  & 0.35(35) \\ 
Aobs 2 & $2.50^{+1.28}_{-0.93}$ & $1.89^{+0.78}_{-0.64}$ & & & 
$0.051\pm0.003$ & 0.30(28) \\ 
Aobs 3 & $1.85^{+1.23}_{-0.84}$ & $1.34^{+0.72}_{-0.59}$ & & & 
$0.063\pm0.004$ & 0.19(28) \\ 
Aobs 4 & $1.87^{+0.05}_{-0.06}$ & $1.35^{+0.04}_{-0.04}$ & 
$6.34^{+0.08}_{-0.08}$ & $70.0\pm15.8$  & $3.49\pm0.02$ & 0.83(324)\\
\enddata
\end{deluxetable}

\clearpage
\begin{deluxetable}{lccccc}
\footnotesize
\tablecaption{RXTE Observation Parameters
\label{robstable}} \tablewidth{0pt}
\tablehead{\colhead{Observation}& \colhead{Obs ID}& \colhead{Date} &
\colhead{Orbital Phase}   & \colhead{On-Time}   & \colhead{Mean
raw PCA rate}\\& &MJD-50400&$T_{\pi/2}=0$&ks&Cts/s } 
\startdata 
Robs 1 &10154-02-01
& 36.40-36.45 & 0.02 & 2.4 & 243 \\ Robs 2 &10154-02-02&
37.39-37.42 & 0.14 & 2.4 & 238 \\ Robs 3 &10154-02-03&
38.39-38.42& 0.25 & 2.4 & 233 \\ Robs 4 &10154-02-04&
39.39-39.42 & 0.37 & 2.4 & 181 \\ Robs 5 &10154-02-05&
40.39-40.42 & 0.49 & 2.4 & 179 \\ Robs 6 & 10154-02-06&
41.05-41.08 & 0.57 & 2.5 & 291 \\ Robs 7 & 10154-02-07 &
42.06-42.12 & 0.70 & 2.6 & 331 \\ Robs 8 & 10154-02-08 &
43.06-43.09 & 0.81 & 2.4 & 333 \\ Robs 9 & 10154-02-09 &
44.05-44.09 & 0.93 & 3.3 & 216 
\enddata
\end{deluxetable}

\begin{deluxetable}{lcccccc}
\footnotesize
\tablecaption{RXTE Spectral Fits from Dec. 1996 Observations \label{spectabler}}
\tablewidth{0pt} \tablehead{\colhead{Observation}& \colhead{Phase} & 
\colhead{$n_H$} & \colhead{$\Gamma$}   & \colhead{Cutoff Energy}
&\colhead{$F_{2-10 keV}$}&\colhead{$\chi^2_{\nu}$}\\
\colhead{} & \colhead{}& \colhead{$\times 10^{22} {\rm cm}^{-2}$} & \colhead{} & \colhead{keV} & \colhead{$\times 10^{-10}$ ergs ${\rm cm}^{-2}
{\rm s}^{-1}$}  &  }
\startdata
Robs1 & 0.995 & $6.31^{+0.43}_{-0.43}$ & $1.251^{+0.029}_{-0.029}$ & 
$13.52^{+0.25}_{-0.25}$ & 2.92 & 0.63 \\
Robs2 & 0.114 & $2.38^{+0.49}_{-0.49}$ & $1.255^{+0.038}_{-0.039}$ &
$14.00^{+0.41}_{-0.40}$ & 1.63 & 0.84 \\
Robs3 & 0.233 & $2.08^{+0.86}_{-0.89}$ & $1.259^{+0.067}_{-0.070}$ &
$13.92^{+1.22}_{-0.87}$ & 0.89 & 0.92 \\
Robs6 & 0.550 & $2.68^{+0.28}_{-0.28}$ & $1.184^{+0.022}_{-0.023}$ &
$13.47^{+0.21}_{-0.22}$ & 3.30 & 1.00 \\
Robs7 & 0.671 & $3.28^{+0.36}_{-0.35}$ & $1.369^{+0.028}_{-0.029}$ &
$13.52^{+0.28}_{-0.29}$ & 4.41 & 1.50 \\
Robs8 & 0.790 & $7.89^{+0.29}_{-0.29}$ & $1.415^{+0.020}_{-0.019}$ &
$13.50^{+0.16}_{-0.17}$ & 4.29 & 0.88 \\
Robs9 & 0.909 & $7.91^{+0.81}_{-0.77}$ & $1.362^{+0.049}_{-0.048}$ &
$13.62^{+0.55}_{-0.49}$ & 1.21 & 0.72 \\
\enddata
\end{deluxetable}


\begin{thebibliography}{}
\bibitem[Benensohn, Lamb, and Taam (1991)]{blt97} Benensohn, J. S., Lamb, D. Q.,
\& Taam, R. E. 1997,\apj, 478, 723
\bibitem[Boroson et al. (1999)]{b99} Boroson, B., Kallman, T., McCray, R., 
Vrtilek, S. D., \& Raymond, J. 1999, \apj, 519, 191 
\bibitem[Brown and Boyle (1984)]{bb84} Brown, J.C. \& Boyle, C.B. 1984, \aap, 141,
369
\bibitem[Castor, Abbott, and Klein (1975)]{cak75} Castor, J. I., Abbott, D. C., \& 
Klein, R. I. 1975, \apj, 195, 157
\bibitem[Chitnis et al. (1993)]{c93} Chitnis, V. R., Rao, A. R., Agrawal, P. C., \& Manchanda, R. K. 1993, \aap, 268, 609
\bibitem[Collura et al. (1987)]{c87} Collura, A., Maggio, A., Sciortino, S.,
Serio, S., Vaiana, G. S., \& Rosner, R. 1987, \apj, 315, 340
\bibitem[Cook and Page (1987)]{cp87} Cook, M. C., \& Page, C.G. 1987,
\mnras, 225, 381
\bibitem[Corbet (1984)]{c84} Corbet, R. 1984, \aap, 141, 91
\bibitem[Corbet (1986)]{c86} Corbet, R. 1986, \mnras, 220, 1047
\bibitem[Cusumano et al. (1998)]{c98} Cusumano, G., Di Salvo, T., Burderi, L., Orlandini, M.,
Piraino, S., Robba, N., \& Santangelo, A. 1998, \aap, 338, L79
\bibitem[Dachs et al. (1986)]{d86} Dachs, J., Hanuschik, R., Kaiser, D.,\&
Rohe, D. 1986, \aap, 159, 276
\bibitem[de Ara\'ujo, de Freitas Pacheco, and Petrini (1994)]{app94}
de Ara\'ujo, F.X., de Freitas Pacheco, J.A., \& Petrini, D. 1994, \mnras, 267, 501
\bibitem[Giacconi et al. (1971)]{g71} Giacconi, R., Kellogg, E.
Gorenstein, P., Gursky, H. \& Tananbaum, H. 1971, \apj, 165, L27
\bibitem[in't Zand, Baykal, and Strohmayer (1998)]{ibs98} in't Zand, J. J. M., Baykal, A., \& Strohmayer, T. E. 1998, \apj, 496,
386
\bibitem[in't Zand, Strohmayer and Baykal (1997)]{isb97} in't Zand, J. J. M., Strohmayer, T. E., \& Baykal, A.  1997, \apj,
479, L47
\bibitem[Iye (1986)]{i86} Iye, M. 1986, \pasj, 38, 463
\bibitem[Koh et al. (1997)]{k97} Koh, D.T., Bildsten, L., Chakrabarty, D., 
Nelson, R.W., Prince, T.A., Vaughan, B.A., Finger, M.H., Wilson, R.B., \&
Rubin, B.C. 1997, \apj, 479, 933
\bibitem[Koyama (1989)]{k89} Koyama, K. 1989, \pasj, 41, 665
\bibitem[Layton et al. (1997)]{l97} Layton, J.T., Blondin, J.M., Owen,
M.P., \& Stevens, I.R. 1997, New Astron., 3, 111
\bibitem[Leahy (1991)]{l91} Leahy, D.A. 1991,\mnras , 250, 310
\bibitem[Leahy and Matsuoka (1990)]{lm90} Leahy, D.A. \& Matsuoka, M. 1990, \apj, 
355, 627
\bibitem[Leitherer (1988)]{l88} Leitherer, C. 1988,\apj, 326, 356
\bibitem[Makishima et al. (1984)]{m84} Makishima, K., Kawai, N., Koyama, K., \& Shibazaki, N.  1984, \pasj,
    36, 679
\bibitem[Makishima et al. (1999)]{mak99} Makishima, K., Mihara, T., Nagase,
F., \& Tanaka, Y. 1999, \apj, 525, 978
\bibitem[Markert et al. (1978)]{m78} Markert, T. H. et al. 1979,
\apjs, 39, 573
\bibitem[Marshall and Rickets (1980)]{mr80} Marshall, N. \& Ricketts, M. J. 1980, \mnras, 193, 7P
\bibitem[Matese and Whitmire (1983)]{mw83} Matese, J. J. \& Whitmire, D. P. 1983, \apj,
266, 776
\bibitem[Mihara (1995)] {m95} Mihara, T. 1995, Ph. D. Thesis, Univ.
of Tokyo
\bibitem[Negueruela et al. (1998)]{n98} Negueruela, I., Reig, P., Coe, M.J.,
\& Fabregat, J. 1998, \aap, 336, 251
\bibitem[Parkes et al. (1980)]{p80} Parkes, G.E., Mason, K.O., Murdin, P.G.,
\& Culhane, J.L. 1980, \mnras, 191, 547
\bibitem[Pravdo et al. (1995)]{pr95} Pravdo, S.H., Day, C.S.R., Angelini, L.,
Harmon, B.A., Yoshida, A., \& Saraawat, P. 1995, \apj, 454, 872
\bibitem[Priedhorsky and Terrell (1984)]{pt84} Priedhorsky, W. C. \&
Terrell, J. 1984, \apj, 280, 661
\bibitem[Ruffert (1999)]{r99} Ruffert, M. 1999, A\&A, 346, 861
\bibitem[Saraswat et al. (1996)]{s96} Saraswat, P., Yoshida, A., Mihara, T.,
Kawai, N., Takeshima, T., Nagase, F., Makishima, K., Tashiro, M.,
Leahy, D.A., Pravdo, S., Day, C.S.R. \& Angelini, L. 1996, \apj, 463, 726
\bibitem[Schwartz et al. (1980)] {s80} Schwartz, D. A., Griffiths,
R. E., Thorstensen, J. R., Charles, P. A. \& Bowyer, S. 1980, \aj,
85, 549
\bibitem[Seward et al. (1976)]{s76} Seward, F. D., Page, C. G.,
Turner, M. J. L., \& Pounds, K. A. 1976, \mnras, 175, 39P
\bibitem[Smith et al. (1985)]{s85} Smith, A., Peacock, A., Jones, L.R. \&
Pye, J.P. 1985, \apj, 296, 469
\bibitem[Underhill and Doazan (1982)]{ud82} Underhill, A.B. \& Doazan, V. ed.
{\it B stars With and Without Emission Lines} 1982, CNRS, Paris and NASA, Washington D.C. (SP-456)
\bibitem[Valinia and Marshall (1998)]{vm98} Valinia, A. \& Marshall, F.E. 1998,
\apj, 505, 134
\bibitem[van Kerkwijk, van Oijen, and van den Heuvel (1989)]{vvv89} van Kerkwijk, M. H., van Oijen, J. G. J., \& van den Heuvel, E. P. J. 1989, \aap, 209, 173
\bibitem[Zamanov et al. (2000)]{z00} Zamanov, R.K., Reig, P., Mart\'i, J.,
Coe, M.J., Fabregat, J., Tomov, N.A., \& Valchev, T. 2000, \aap  accepted, 
astro-ph/0012371
\end{thebibliography}
\end{document}